\documentclass[amsmath,nobibnotes,aps,superscriptadress,amssymb,pra,aps,showpacs,superscriptaddress,twocolumn]{revtex4-1}
\usepackage{amsmath,amsfonts,amssymb,amsthm,graphics,graphicx,epsfig,bbm}
\usepackage[colorlinks=true,citecolor=blue,linkcolor=blue,urlcolor=blue]{hyperref}
\usepackage[usenames]{color}
\usepackage{graphicx}
\usepackage{amsmath}
\usepackage{epsfig}
\usepackage{dcolumn}
\usepackage{bm}
\usepackage{color}
\usepackage{times}
\usepackage{epstopdf}
\usepackage{amssymb}
\usepackage{amstext}
\usepackage{latexsym}
\usepackage{hyperref}
\usepackage{amsfonts}
\usepackage{psfrag}
\usepackage{soul,xcolor}
\usepackage[normalem]{ulem}
\usepackage{dsfont}
\usepackage{physics}
\usepackage{footnote}
\usepackage{tikz}
\usepackage{blindtext}
\usepackage{bbm}
\usepackage{float}

\renewcommand{\Re}{\mathrm{Re}}

\usepackage{algorithm}
\usepackage{algpseudocode}
\makeatletter
\newcommand{\StatexIndent}[1][3]{%
  \setlength\@tempdima{\algorithmicindent}%
  \Statex\hskip\dimexpr#1\@tempdima\relax}
\makeatother

\begin{document}

\title{Quantum Genetic Algorithm with Individuals in Multiple Registers}

\author{Rub\'en Ibarrondo}\email{ruben.ibarrondo@ehu.eus} 
\affiliation{Department of Physical Chemistry, University of the Basque Country UPV/EHU, Apartado 644, 48080 Bilbao, Spain}
\affiliation{EHU Quantum Center, University of the Basque Country UPV/EHU, Bilbao, Spain}

\author{Giancarlo Gatti}
\affiliation{Department of Physical Chemistry, University of the Basque Country UPV/EHU, Apartado 644, 48080 Bilbao, Spain}
\affiliation{EHU Quantum Center, University of the Basque Country UPV/EHU, Bilbao, Spain}
\affiliation{Quantum Mads, Uribitarte Kalea 6, 48001 Bilbao, Spain}

\author{Mikel Sanz}\email{mikel.sanz@ehu.es} 
\affiliation{Department of Physical Chemistry, University of the Basque Country UPV/EHU, Apartado 644, 48080 Bilbao, Spain}
\affiliation{EHU Quantum Center, University of the Basque Country UPV/EHU, Bilbao, Spain}
\affiliation{Ikerbasque Foundation for Science, Plaza Euskadi 5, 48009 Bilbao, Spain}
\affiliation{BCAM-Basque Center for Applied Mathematics, Mazarredo, 14, 48009 Bilbao, Spain}

\begin{abstract} 

Genetic algorithms are heuristic optimization techniques inspired by Darwinian evolution, which are characterized by successfully finding robust solutions for optimization problems. Here, we propose a subroutine-based quantum genetic algorithm with individuals codified in independent registers. This distinctive codification allows our proposal to depict all the fundamental elements characterizing genetic algorithms, i.e. population-based search with selection of many individuals, crossover, and mutation. Our subroutine-based construction permits us to consider several variants of the algorithm. For instance, we firstly analyze the performance of two different quantum cloning machines, a key component of the crossover subroutine. Indeed, we study two paradigmatic examples, namely, the biomimetic cloning of quantum observables and the Bu\v zek-Hillery universal quantum cloning machine, observing a faster average convergence of the former, but better final populations of the latter. Additionally, we analyzed the effect of introducing a mutation subroutine, concluding a minor impact on the average performance. Furthermore, we introduce a quantum channel analysis to prove the exponential convergence of our algorithm and even predict its convergence-ratio. This tool could be extended to formally prove results on the convergence of general non-unitary iteration-based algorithms.

\end{abstract}

\maketitle

\section{Introduction}

Genetic Algorithms (GAs) are bioinspired algorithms with well established performance in finding resilient solutions to complex optimization problems, such as problems with exponentially large search spaces and noisy optimization criteria \cite{Hornby2006, Holland1992, DeJong1993, Zebulum2001}. In these algorithms, every element of the search space can be potentially represented by an individual and the selection towards the optimal solution is performed by Darwinian-like evolution. The set of individuals is known as population, their codification is commonly known as chromosomes, and their performance is ranked by a fitness function. Although there is no formal definition of GA which univocally distinguishes them from other evolutionary algorithms, there is a general consensus about the presence of the following four characteristic elements: population-based search through joint evolution of a set of individuals, a selection of some of them according to their performance, a crossover operation to breed new individuals, and a mutation operation which randomly modifies them \cite{Melanie1996}.

Optimization problems are cornerstone in real-life applications, and consequently huge efforts have been devoted to the development of quantum algorithms for certain approximate optimization problems in which the results proving that finding an approximation with sufficiently small error bound is NP-complete are not applicable. In this cases, there is a chance for exponential speedup by quantum computation. Nonetheless, even potential improvements in a factor or exponent could be of relevance in comercial and industrial applications and it encourages us to carry on with the research in alternative heuristic quantum approaches. During the past decades, the merge of GAs and quantum computation has been a source of new heuristic optimization methods \cite{Sofge2008, Roy2014, Lahoz-Beltra2016}. Induced by the non-linear behavior of genetic operators, most of the effort has been focused on quantum inspired GAs, which integrate some concepts of quantum mechanics to engineer new varieties of classical evolutionary algorithms \cite{Narayanan1996, Han2000, Han2001, Han2002, Yang2004, Wang2005, Yingchareonthawornchai2012, Roy2014}. On the other hand, fully quantum approaches potentially achieving quantum speed-up have only attained partial success in the inclusion of the aforementioned characteristic elements \cite{Rylander2001, Udrescu2006, Malossini2008, Saitoh2014}. In 2001, Rylander et al. proposed in Ref. \cite{Rylander2001} a quantum genetic algorithm introducing the concepts of chromosome-register and fitness-register. The individuals were encoded in the chromosome-registers which are possibly entangled with fitness-registers due to quantum superposition. Although it was claimed that quantum superposition provides an increased searching power, this conclusion has been considered unsupported due to the lack of heuristic or analytic evidence \cite{Sofge2008}. In 2006, Udrescu et al. proposed in Ref. \cite{Udrescu2006} an algorithm based on quantum searching and inspired by evolutionary computation, called reduced quantum GA. Here, individuals are represented by a binary basis so that the whole population can be encoded as a quantum superposition in a single register. This encoding is compatible with Grover's quantum search algorithms which leads to a speed-up in the selection process \cite{Grover1996}. However, they conclude that, for a non-structured search, there is no need for elements like crossover or mutation, which raises the question of whether it should be considered a GA \cite{Sofge2008}. Afterwards, Malossini et al. \cite{Malossini2008} proposed in 2008 the quantum genetic optimization algorithm, which also employed quantum searching techniques. These techniques enhance the selection procedure similarly to the previous work but the crossover and mutation subroutines are still classically introduced. Lately, Saitoh et al.~\cite{Saitoh2014} extended both previous proposals by introducing an algorithm which includes crossover and mutation as quantum subroutines. However, the treatment of the population substantially differs from its role in classical GAs. Indeed, as the selection procedure is implemented by projective measurements, only one individual is selected in each generation, lacking the population-based search feature, and consequently, reducing the intrinsic exploration capacity with respect to GAs.

In this article, we propose a quantum genetic algorithm (QGA) with individuals codified in independent registers, allowing for population-based search and selection, which are characterizing elements of GAs and a distinctive feature with respect to previous approaches. Our proposal is composed of modular quantum subroutines inspired by classical GAs: selection, crossover, and mutation. The intrinsic non-linear nature of selection and crossover leads to fundamental obstacles posed by the principles of quantum information, such as no-cloning and no-deleting theorems. Selection is performed by a reversible Hamiltonian-based sorting with ancillary qubits and a posterior partial trace of the lowest ranked individuals. The replication step of the crossover is carried out by a partial quantum cloning machine and the combination step is accomplished by swap gates. This modular structure of the subroutines allows us to benchmark two paradigmatic quantum approximated cloning machines: the biomimetic cloning of quantum observables and the Bu\v zek-Hillery universal quantum cloning machine. Numerical analysis shows a faster average convergence of the former, but better final populations of the latter. Lastly, mutations are introduced by randomly allocated Pauli gates. We conclude that their presence has a negligible effect upon the average performance. Finally, we present a toolbox based on the spectral theory of quantum channels, which we employ to formally prove the exponential convergence of the algorithm, as well as to predict its convergence rate. Remarkably, both this prediction and the final quantum state accurately match our numerical simulations. This approach can be extended to other non-unitary iteration-based quantum algorithms.

Section~\ref{sec:qga} introduces the QGA and its constituent subroutines, and is concluded by an analysis of the convergence. Section~\ref{sec:results} presents the main results regarding convergence and performance of the variants.

\section{Quantum genetic algorithm (QGA)}
\label{sec:qga}

In order to take advantage of the exploratory capacity of population-based search, we encode individuals in several independent quantum registers. This allows to select and replicate them on a population level. More specifically, let us consider $n$ individuals comprised of $c$-qubits each, which we assume for simplicity to be divisible by $4$ and $2$, respectively. The search space is the Hilbert space associated to a $c$-qubit quantum register, $\mathcal{H}$. We rate the individuals by the average energy of their state according to a problem Hamiltonian, $H_P$, which describes a cost function to be minimized.

As previously mentioned, the present QGA is comprised of selection, crossover and mutation subroutines. Selection is performed by sorting the population and then discarding the worst individuals given by $H_P$. Crossover is performed by replicating the selected individuals employing an approximate quantum cloning machine (QCM) \cite{Scarani2005} and then combining their features performing qubit swap operations. Finally, mutation is performed applying single qubit rotations at random. The discard and copy steps in these subroutines have required approximate implementations, due to fundamental limitations imposed by the no-deleting and no-cloning theorems \cite{Pati2000, Wootters1982}. The modular flexibility of our proposal allows us to analyze four variants given by suppressing or activating the mutation, as well as the use of two different QCMs, namely, the biomimetic cloning of quantum observable (BCQO) and the Bu\v zek-Hillery universal quantum cloning machine (UQCM). The full QGA implementation is summarized in Algorithm~\ref{alg:qga}, and each subroutine is detailed in the following subsections.

Additionally, we analyze the algorithm in terms of quantum channels. This approach is based on the operator sum representation  of each of the subroutines of the QGA process, i.e. a general description of a quantum state transformation. This is given by $\mathcal{T}(\rho) = \sum_k E_k \rho E_k^{\dag}$, where $E_k$ are Kraus operators, and $\rho$ an arbitrary density matrix. This way, each iteration is also described by a quantum channel and the full QGA process corresponds to its self-composition. Consequently, we can employ the spectral theory of quantum channels to formally prove an exponential convergence of the algorithm, as well as bound its convergence rate.

\begin{algorithm}[H]
\caption{Quantum genetic algorithm}\label{alg:qga}
\begin{algorithmic}[1]
\State $n \gets number\_of\_registers$ \Comment{assumed divisible by four}
\State $c \gets number\_of\_qubits\_per\_register$ \Comment{assumed even}
\State Initialize with a random state
\Repeat
	\State \textbf{sort} registers $1$ to $n$
	\State \textbf{reset} registers $n/2$ to $n$
	\For{ $r=1,2,\dots, n/2$}
		\State \textbf{pseudo-clone} register $r$ to register $n/2+r$.
	\EndFor
	\For{ $i=0,1,\dots, n/4-1$}
		\State \textbf{swap} the last $c/2$ qubits of register $n/2 + 2 i + 1$
	    \StatexIndent[2] with the last $c/2$ qubits of register $n/2 + 2 i + 2$.
	\EndFor
	\State \textbf{mutate} each qubit with probability $p_m$
 \Until{ending criteria is met $\lor$ $G$ generations} 
\end{algorithmic}
\end{algorithm}

\subsection{Quantum Selection Subroutine}
\label{sec:qselection}
%
%\begin{enumerate}
%	\item QSS is implemented in two steps.
%	\item Sorting network.
%	\item Sorting oracle. Comments on pairwise sort.
%	\item Discarding. Concerns about no-cloning theorem and the risk to lose entangled information in upper-registers.
%	\item Comments on the final state in terms of the eigenvectors of $H_P$.
%	\item Comment on realistic implementation of the pairwise sort. 
%\end{enumerate}

\begin{figure*}[!hbt]
\centering
\includegraphics[width=0.65\textwidth]{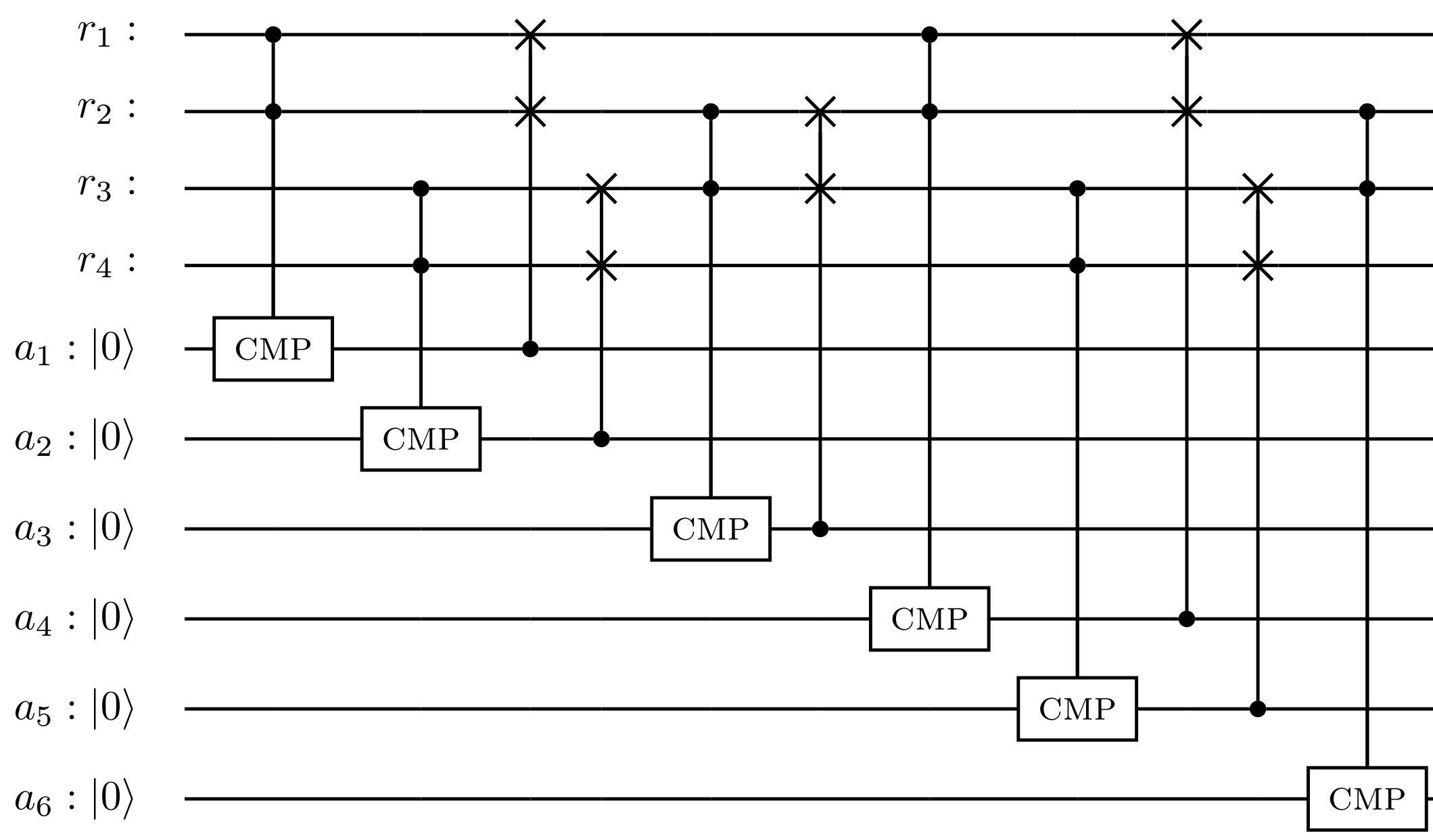}
%\begin{quantikz}[column sep=0.25cm, row sep=0.3cm]
%\lstick{$r_1:$} & & \ctrl{4}  & \qw  &  \swap{1} & \qw & \qw & \qw & \ctrl{7} & \qw & \swap{1} & \qw & \qw & \qw & \qw \\
%%
%\lstick{$r_2:$} & & \ctrl{3}  & \qw & \swap{} & \qw & \ctrl{5} & \swap{1} & \ctrl{6} & \qw & \swap{} & \qw & \ctrl{8} & \swap{1} & \qw\\
%%
%\lstick{$r_3:$} & & \qw & \ctrl{3} & \qw & \swap{1} & \ctrl{4} & \swap{} & \qw & \ctrl{6} & \qw & \swap{1} & \ctrl{7} & \swap{} & \qw \\
%%
%\lstick{$r_4:$} & & \qw  & \ctrl{2} & \qw & \swap{} & \qw & \qw & \qw & \ctrl{5} & \qw & \swap{} & \qw & \qw & \qw \\
%%
%\lstick{$a_1: \ket{0} $} & & \gate[wires=1][1cm]{\textsc{cmp}} & \qw & \ctrl{-4} & \qw & \qw & \qw & \qw & \qw & \qw & \qw & \qw & \qw & \qw\\
%%
%\lstick{$a_2: \ket{0}$} & & \qw &  \gate[wires=1][1cm]{\textsc{cmp}}  & \qw & \ctrl{-3} & \qw & \qw & \qw & \qw & \qw & \qw & \qw & \qw & \qw\\
%%
%\lstick{$a_3: \ket{0}$} & & \qw &  \qw & \qw & \qw & \gate[wires=1][1cm]{\textsc{cmp}} & \ctrl{-5} & \qw & \qw & \qw & \qw & \qw & \qw & \qw \\
%%
%\lstick{$a_4: \ket{0}$} & & \qw &  \qw & \qw & \qw & \qw & \qw & \gate[wires=1][1cm]{\textsc{cmp}} & \qw & \ctrl{-7} & \qw & \qw & \qw & \qw \\
%%
%\lstick{$a_5: \ket{0}$} & & \qw &  \qw & \qw & \qw & \qw & \qw & \qw & \gate[wires=1][1cm]{\textsc{cmp}} & \qw & \ctrl{-6} & \qw & \qw & \qw\\
%%
%\lstick{$a_6: \ket{0}$} & & \qw &  \qw & \qw & \qw & \qw & \qw & \qw & \qw & \qw & \qw & \gate[wires=1][1cm]{\textsc{cmp}} & \ctrl{-8} & \qw
%\end{quantikz}
\caption{Circuit representation for the sorting network with 4 registers. The controlled \textsc{cmp} denotes the operation of $O_{\textsc{cmp}}$ on the ancillary qubits defined in Eq.~\ref{eq:sorting_oracle}. The registers $r_i$ initially store the states of the unsorted population, $\sum_{\bm{k}}b_{\bm{k}}\ket{u_{\bm{k}}}$, and $a_i$ are the ancillary qubits required to store the permutations performed, which ensure reversibility of the circuit. The output state is a superposition of sorted populations possibly entangled with the ancillas, $\sum_{\bm{k}} b_{\bm{k}} \ket{u_{s(\bm{k})}}\ket{\sigma(\bm{k})}$, as shown in Eq.~\ref{eq:sorted_state}.}
\label{fig:sorting_circuit}
\end{figure*}

The quantum selection subroutine aims to select the best individuals of the population without measuring their state. This subroutine consists in sorting the individuals to discard the lower half of the population. We refer to registers $r_1$ to $r_{n/2}$  and $r_{n/2+1}$ to $r_n$ as the upper-registers and lower-registers, respectively.

Sorting networks are protocols that sort the joint state of an $n$-register system~\cite{cormen01}. We use the Bubble Sort algorithm, which is not asymptotically optimal but performs efficiently for a moderate number of registers. For $n$-individual populations, this algorithm is composed of $n$ layers. The odd ones sort consecutive register pairs starting from the first register performing $n/2$ comparisons, whereas the even layers start from the second register and perform $n/2-1$ comparisons, hence the total number of required comparisons is $\frac{n(n-1)}{2}$. Analogously, quantum sorting networks substitute the classical pairwise comparison by a quantum pairwise sorting operator \cite{Berry2018}. We propose a quantum sorting network implementation tailored to our algorithm in the following.

Let us first describe the states in terms of the problem basis for $\mathcal{H}$, which is the sorted $H_P$ eigenbasis  $\qty(\ket{u_1}, \dots, \ket{u_{2^c}})$,  in increasing eigenvalue order, $\epsilon_{k}\leq \epsilon_{k+1}$, where $H_P\ket{u_k}=\epsilon_k\ket{u_k}$. Consequently, the problem basis for the Hilbert space $\mathcal{H}^{\otimes n}$ of an $n$-individual population is formed by
\begin{equation}
\ket{u_{\bm{k}}} = \ket{u_{k_1}}_1 \otimes \cdots \otimes \ket{u_{k_n}}_n \text{,}
\end{equation}
with  $k_i \in \qty{1, \dots, 2^c}$. We refer to $\bm{k}=(k_1, \dots, k_n)$ as the population index sequence which labels the population basis state $\ket{u_{\bm{k}}}$.

The pairwise sorting operator is defined by concatenating a comparison oracle $O_{\textsc{cmp}}$ with a controlled swap $C_{\textsc{swap}}$ acting on consecutive registers $\{i,i+1\}$ and ancillary qubit $a$,
\begin{multline}
O_{\textsc{cmp}} \ket{u_k}_i\ket{u_{k'}}_{i+1}\ket{0}_a = \\
\begin{cases}
\ket{u_k}_i\ket{u_{k'}}_{i+1}\ket{0}_a & \text{if } \epsilon_{k'} \leq \epsilon_{k}\\
\ket{u_k}_i\ket{u_{k'}}_{i+1}\ket{1}_a & \text{if } \epsilon_{k'} > \epsilon_{k}\\
\end{cases}
\label{eq:sorting_oracle}
\end{multline}
and
\begin{equation}
C_{\textsc{swap}} \ket{x}_i\ket{y}_{i+1}\ket{c}_a = 
\begin{cases}
\ket{x}_i\ket{y}_{i+1}\ket{c}_a & \text{if } c = 0\\
\ket{y}_i\ket{x}_{i+1}\ket{c}_a & \text{if } c = 1
\end{cases}\text{,}
\label{eq:cswap}
\end{equation}
where $\ket{x}\ket{y}$ represents any separable pure state. Figure~\ref{fig:sorting_circuit} shows the circuit representation for a four-register quantum sorting network, where $O_{\textsc{cmp}}$ is denoted by controlled \textsc{cmp} gates. For the purpose of clarity, let us note that $O_{\textsc{cmp}}$ can indeed be implemented without explicitly knowing the eigenvalues and eigenstates of the problem Hamiltonian. On a general basis, in order to compare two individuals two additional registers can be included to compute their energy in binary representation and mark the ancillary qubit with an arithmetic comparator~\cite{Cuccaro2004}. For instance, a quantum phase estimation subroutine can be employed to compute the energies, which only relies on the ability to perform controlled $e^{\pm i \theta H_P}$ gates~\cite{nielsen00}.

For terminology simplicity, we define $s(\bm{k})$ as an ascending permutation of the population index sequence $\bm{k}$, i.e. $s(\bm{k})_i \leq s(\bm{k})_{i+1}$. This allows us to define $\ket{u_{s(\bm{k})}}$ as a basis state of the population with sorted individuals. We also define $\sigma(\bm{k})$ as the binary sequence of instructions associated to a given input and sorting network, e.g. $\frac{n(n-1)}{2}$ individual swap instructions given by a Bubble Sort. In the quantum circuit implementation, $\sigma(\bm{k})$ corresponds to the outputs of the ancillas given population input $\ket{u_{\bm{k}}}$. By convention we assume no permutations if the index sequence is already sorted, i.e. $\sigma(\bm{k})=(0,\dots,0)$.

In general, given initial state $\sum_{\bm{k}} b_{\bm{k}} \ket{u_{\bm{k}}}$, the sorting output is a superposition of sorted populations, possibly entangled with the ancillary qubits,
\begin{equation}\label{eq:sorted_state}
\ket{\Psi} = \sum_{\bm{k}} b_{\bm{k}} \ket{u_{s(\bm{k})}}\ket{\sigma(\bm{k})}\text{.}
\end{equation}
After the sorting, ancillary qubits must be discarded in order to proceed with the algorithm, therefore, the output state is a reduced density matrix of $\ketbra{\Psi}{\Psi}$,
\begin{equation}
\rho_{\text{sorted}} = \sum_{\bm{k}} \sum_{\bm{k}'} b_{\bm{k}} b_{\bm{k}'}^* \delta_{\sigma(\bm{k}), \sigma(\bm{k}')} \ketbra{u_{s(\bm{k})}}{u_{s(\bm{k}')}}\text{,}
\end{equation}
where $\delta_{i, j}$ is the Kronecker delta. The state $\rho_{\text{sorted}}$ is a mixture of pure states, each being a quantum superposition of states with equal sorting instructions, $\sigma(\bm{k}) = \sigma(\bm{k}')$. Hence, the Kraus operators of this subroutine are
\begin{equation}
A_{\kappa} = \sum_{\bm{k}} \delta_{\sigma(\bm{k}), \kappa} \ketbra{u_{s(\bm{k})}}{u_{\bm{k}}}\text{,}
\end{equation}
which transform the states onto their sorted versions if they are sorted by a given set of permutation instructions $\kappa$. Here, $\kappa$ represents an arbitrary sorting instruction bitstring and $\bm{k}$ iterates over all the possible index sequences.

Finally, we discard the state in the lower-registers and replace it with a reference state $\rho_0^{\otimes n/2}$, obtaining
\begin{equation}\label{eq:reset}
\rho_{\text{selected}} = \tr_{\text{low}}(\rho_{\text{sorted}}) \otimes \rho_0 ^{\otimes n/2}\text{.}
\end{equation}
We refer to this process as the reset of the lower-registers. Given a reference state with spectral decomposition $\rho_0=\sum_{r=1}^d \omega_r \ketbra{e_r}$ and $d\leq 2^c$, the Kraus operators of the reset step are
\begin{equation}
B_{j,r_1,\dots,r_{n/2}} = \sqrt{\omega_{r_1}\dots \omega_{r_{n/2}}}\; \mathbb{I}^{\otimes \frac{n c}{2}} \otimes \ketbra{e_{r_1}\dots e_{r_{n/2}}}{j}\text{,}
\end{equation}
where the states $\ket{j}$ form an orthogonal basis of the state space of the lower-registers, with $j=1,\dots,2^{\frac{n c}{2}}$. Note that each Kraus operator corresponds to one classical statistical outcome of the lower-register reset. The value of $\rho_0$ is chosen to correspond to the crossover subroutine, explained in the next section, and is dependent on the cloning machine used. For instance, for BCQO the reference state is a pure state, $\rho_0=\ketbra{e_0}$, hence, $B_j= \mathbb{I}^{\otimes n c/2} \otimes \ketbra{e_0\dots e_0}{j}$. Whereas for UQCM $\rho_0=\mathbb{I}/2^c$, therefore $B_{j,r}=\sqrt{\frac{1}{2^{n c/2}}} \mathbb{I}^{\otimes n c/2} \otimes \ketbra{r}{j}$, where the states $\ket{r}$ form an orthogonal basis of the state space of the lower-registers too, $r=1,\dots,2^{\frac{nc}{2}}$.

The proposed subroutine provides a population-based approach to sort a set of individuals with unitary-preserving rules, as well as discard the lower-registers with the same fundamental limitations. This allows us to implement the selection step in an optimization process defined by the problem Hamiltonian in a quantum-compatible manner.

\subsection{Quantum Crossover Subroutine}

\begin{figure*}[tb]
\centering
\includegraphics[scale=0.52]{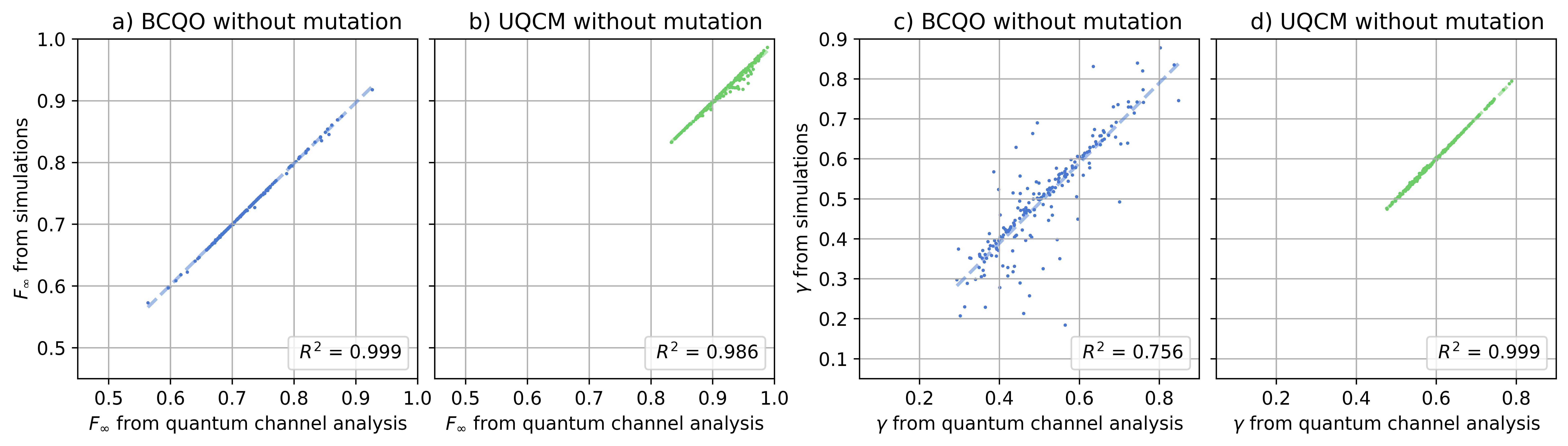}
\caption{Comparison of results obtained from numerical simulations and quantum channel analysis for a sample of problem Hamiltonians. Figures (a) and (b) represent the final fidelity $F_{\infty}$ between the best individual and the target ground state obtained respectively for BCQO and UQCM variants, both without mutation. Figures (c) and (d) represent the fidelity convergence rate $\gamma$ obtained respectively for BCQO and UQCM variants, both without mutation. The large lineal correlation $R^2$ is consistent with both results being the same in all cases.}
\label{fig:qch_vs_sim}\label{fig:qch_vs_sim_a}\label{fig:qch_vs_sim_b}\label{fig:qch_vs_sim_c}\label{fig:qch_vs_sim_d}
\end{figure*}

In this subroutine the states of the selected individuals in the upper-registers are approximately replicated into the lower-registers by means of quantum cloning machines (QCMs). More specifically, each QCM acts on the joint state of registers $i$ and $\frac{n}{2}+i$, where $1\leq i\leq\frac{n}{2}$. Afterwards, consecutive lower-registers are combined swapping their second half.

We denote the QCM acting on individuals $i$ and $j$ by $T_{\textsc{qcm}}^{i, j}$. The global replication step is described by the operation
\begin{equation}\label{eq:global_replication}
\rho_{\text{replicated}} = 
\left[\bigotimes_{i=1}^{n/2} T_{\textsc{qcm}}^{i, i+n/2}\right] (\rho_{\text{selected}})\text{.}
\end{equation}
If we represent QCM $\mathcal{T}_{\text{QCM}}^{i,j}$ with Kraus operators $c_{k}^{i,j}$ for $k=1,\dots,K$, the Kraus operators for the global replication are 
\begin{equation}
C_{k_1, \dots, k_{n/2}} = \bigotimes_{i=1}^{n/2} c_{k_i}^{i, i+n/2}\text{,}
\end{equation}
where each element of the sequence $k_1, \dots, k_{n/2}$ takes values from 1 to $K$.

Finally, we swap qubits $\frac{c}{2}$ to $c$ in register $\frac{n}{2}+2i - 1$, with qubits $\frac{c}{2}$ to $c$ in register $\frac{n}{2}+2i$ for all $1\le i \le\frac{n}{4}$, which yields
\begin{equation}\label{eq:uswap}
U_{\text{swap}} = 
\prod_{i=nc/4}^{(n-2)c/2}
\prod_{j=\frac{c}{2}+1}^{c}
  S_{2i+j, 2i+c+j}\text{,}
\end{equation}
where $S_{i,j}$ denotes a swap between qubits $i$ and $j$, and qubits are numbered increasing from the first one in $r_1$ to the last one in $r_n$. Recall that $c$ is the qubit distance between consecutive individuals. The population after crossover is given by
\begin{equation}
\rho_{\text{crossover}} = 
U_{\text{swap}}\qty(\left[\bigotimes_{i=1}^{n/2} T_{\textsc{qcm}}^{i, i+n/2}\right] (\rho_{\text{selected}})) U_{\text{swap}}^{\dag}\text{.}
\end{equation}

\subsubsection{Quantum cloning machines}\label{sec:qcm}

Generically, let us consider a joint system $A$--$B$ such that $A$ is initially in an unknown state $\ket{\psi}$ and $B$ in a reference state $\ket{R}$. Then, a quantum cloning machine (QCM) is a quantum operation mapping $\ket{\psi}_A\ket{R}_B$ to an output state $\rho_{AB}'$ which contains two approximated copies of the input state $\ket{\psi}$~\cite{Scarani2005}. The quality of the copies is measured in terms of the fidelity of $\ket{\psi}$ with the partial states of $\rho_{AB}'$ of each clone, i.e
\begin{equation}
F_j(\ket{\psi}) = \expval{\rho_j'}{\psi}\text{,} \quad j=A, B
\end{equation}
where $\rho_A' = \tr_B[\rho_{AB}']$ and $\rho_B' = \tr_A[\rho_{AB}']$, which is known as the singe copy fidelity. %According to the classification provided in Ref.~\cite{Scarani2005}, we focus on symmetric QCMs, which satisfy $F_A=F_B$, and QCMs producing two approximated copies from a single one.

Formally, let $\mathcal{H}_A$ and $\mathcal{H}_B$ be the $d$-dimensional Hilbert spaces associated to systems $A$ and $B$, respectively. In general, a QCM is described as a quantum operation $T_{\textsc{qcm}} : \mathcal{H}_A\otimes \mathcal{H}_B \rightarrow \mathcal{H}_A \otimes \mathcal{H}_B$. However, the action of $T_{\textsc{qcm}}$ on an input state $\rho_{AB} \in \mathcal{H}_{A} \otimes \mathcal{H}_{B}$ is only considered an approximated cloning operation if $\rho_{AB} = \rho_A \otimes \rho_0$, where $\rho_A \in \mathcal{H}_A$ is the arbitrary state to be copied and $\rho_0 \in \mathcal{H}_B$ is the reference state to be overwritten.

Below, we analyze the performance of our algorithm employing two possible QCMs. Namely, the biomimetic cloning of quantum observables (BCQO)~\cite{ARodriguez2014} and the optimal symmetric universal quantum cloning machine (UQCM) introduced by Bu\v zek and Hillery~\cite{Buzek1996, Werner1998}.

\paragraph{BCQO.} 

For a reference state $\rho_0$ on system $B$ and a quantum observable $\theta$, the cloning operator $U(\theta, \rho_0)$ is the operator satisfying
\begin{equation}\label{eq:biomimetic_condition}
\expval{\theta}_{\rho} \equiv 
\tr[\theta\rho] = 
\expval{\theta\otimes \mathbb{I} }_{U \rho\otimes \rho_0 U^{\dag}} =  
\expval{\mathbb{I} \otimes \theta}_{U \rho\otimes \rho_0 U^{\dag}}\text{,}
\end{equation}
for any state $\rho$ on system $A$. That is, the expected value of $\theta$ on the initial state of system $A$ is the same as the expected value that is eventually obtained on both systems. If $U$ satisfies Eq.~\ref{eq:biomimetic_condition} for the operator $\theta$, then it also holds for a complete set of observables commuting with $\theta$. Additionally, the unitary operator can be straightforwardly described in the $\qty{\ket{j}}_{j=1}^n$ basis which diagonalizes those observables, assuming the reference state is $\rho_0 = \ketbra{1}$\cite{ARodriguez2014}.

%The single copy fidelity is equal for both systems and depends on the cloned state. For instance, the single copy fidelity for a pure state $\ket{\psi} = \sum_{j=1}^n a_j \ket{j}$ is
%\begin{equation}\label{eq:fidelity_bcqo}
%F(\ket{\psi}) = \sum_{j=1}^n \abs{a_j}^4.
%\end{equation}
%As a matter of choice, we employ a BCQO associated to a $\theta$ diagonal in the computational basis. %Consequently, the information of the statistics of the computational basis is perfectly replicated and preserved. Whereas, other quantum information is spread onto the population and it is susceptible of disappearing after the selection process.

\paragraph{UQCM.}

Let $\mathcal{H}_+^2$ be the subspace of $\mathcal{H}_{A} \otimes \mathcal{H}_{B}$ formed by the invariant states with respect to  $A$--$B$ system swap $S_{A,B}$, and let $S_+$ be the projection operation into $\mathcal{H}_+^2$,
\begin{equation}
S_+ \equiv \frac{1}{2}\qty(\mathbb{I} _A\otimes \mathbb{I}_B + S_{A,B})\text{.}
\end{equation}
We require the reference state $\rho_0 = \frac{\mathbb{I}}{d}$. Recall that $\mathcal{H}_A$ and $\mathcal{H}_B$ are $d$-dimensional spaces. Then, given a state to copy $\rho_A$, the UQCM is performed by projecting the state $\rho_A \otimes \rho_0$ into the symmetric subspace $\mathcal{H}_+^2$, and normalizing the result, i.e.
\begin{equation}
T_{\textsc{uqcm}}(\rho) = \frac{2 d}{d+1} S_+ \qty(\tr_{B}(\rho) \otimes \rho_0) S_+\text{,}
\end{equation}
where we require the partial trace on $B$ so that the operation is well defined for every possible input state $\rho$. However, the operation is only considered an approximated cloning for separable cases $\rho = \rho_A \otimes \rho_0$, which satisfy $\tr_B(\rho)=\rho_A$. The Kraus operators are
\begin{equation}
c_{r,k}^{i,j} = \sqrt{\frac{2}{2^c+1}} S_+^{i,j} \qty(\mathbb{I}^i \otimes \ketbra{r}{k}^j)\text{,}
\end{equation}
where the states $\ket{r}$ and $\ket{k}$ form an orthogonal basis of the $j$th register state space and $S_+^{i,j}$ represents the projection into the swap-invariant subspace of registers $i$ and $j$.

%The reduced states have the form
%\begin{equation}
%\rho_A'=\rho_B' = \eta \rho_A + \qty(1-\eta) \rho_0,
%\quad \text{with} \quad
%\eta = \frac{1}{2} \frac{2+d}{1+d}.
%\end{equation}
%Therefore, the single copy fidelity of the UQCM is
%\begin{equation}\label{eq:fidelity_uqcm}
%F = \frac{1}{2} + \frac{1}{1+d},
%\end{equation}
%which is clearly independent of the input state $\rho$.
      
\subsection{Quantum Mutation subroutine}
\label{sec:qmutation}

The aim of mutation is to slightly perturb some of the individuals, in order to facilitate the exploration of new areas in the search space. Classically, it is often implemented by bit-flip operations according to a sufficiently small mutation probability. Similarly, its quantum analogue is implemented by mutation unitary gates which are applied stochastically on each individual. In order to express mutation in terms of Kraus operators, let us consider a generic set of mutation gates $\qty{U_{\mu}}_{\mu=0}^M$ with disjoint probabilities $\qty{p_{\mu}}_{\mu=0}^M$, defining $U_{0}=\mathbb{I}$ for the zero-mutation case. Then, the Kraus operators are all possible gate combinations
\begin{equation}
D_{\mu_1, \dots, \mu_n} = \sqrt{p_{\mu_1}\cdots p_{\mu_n}}\; U_{\mu_1}\otimes \cdots\otimes U_{\mu_n}\text{.}
\end{equation}

The most suitable set of mutation gates may depend on the characteristics of the considered problem and exploring large search spaces could require elaborated approaches, such as including two-qubit gates or generalized rotations. Here, we choose an ensemble of gates composed by single-qubit mutations, namely the Pauli gates $X$, $Y$ and $Z$. More precisely, each qubit is mutated with probability $p_m$, and the mutation gate is chosen at random from $\{X,Y,Z\}$ with equal probability. Hence, the mutation gate ensemble is $U_{\mu} \in \qty{\mathbb{I}, X, Y, Z}^{\otimes c}$ with $p_{\mu} = \qty(\frac{p_m}{3})^k(1-p_m)^{c-k}$, where $k$ is the number of non-identity gates in $U_{\mu}$. 

\subsection{Fixed points and convergence}
\label{sec:fp_convergence}

Let us call $\mathcal{T}$ the population evolution for a single generation and $\rho_\text{in}$ the initial state for the population. The operation $\mathcal{T}$ is the composition of the subroutines described in previous subsections, namely, reset $\mathcal{T}_R$, crossover $\mathcal{T}_C$, swap $U_{\text{SWAP}}$, mutation $\mathcal{T}_M$, and sorting $\mathcal{T}_S$ quantum channels, and it can be expressed as  
\begin{equation}
\mathcal{T}=\mathcal{T}_S \mathcal{T}_M U_{\text{SWAP}} \mathcal{T}_C \mathcal{T}_R \text{,}
\end{equation}
which is also a quantum channel. Consequently, the outcome after $G$ generations of the QGA corresponds to $G$ applications of $\mathcal{T}$ to the initial state, $\rho(G;\rho_\text{in})=\mathcal{T}^G(\rho_\text{in})$. Therefore, the performance and convergence of the algorithm are given by the algebraic properties of the quantum channel in its asymptotic limit, i.e. the convex set of fixed points and the spectral subradius. In order to analyze these properties, let us introduce the eigenvectors and eigenvalues of $\mathcal{T}$,
\begin{equation}
\label{eq:eigenvalueeq}
\mathcal{T}(W_l)=\lambda_l W_l,\quad \text{with}\; \abs{\lambda_{l+1}}\leq \abs{\lambda_l}\text{.}
\end{equation}
As $\mathcal{T}$ is a completely positive trace preserving map, according to the Perron-Frobenius Theorem the spectral radius is one \cite{Sanz2010}. The convex set spanned by the eigenvectors corresponding to the eigenvalues with absolute value one is called the convex set of fixed points. Moreover, we assume w.l.o.g. that the first $m\geq 1$ eigenvalues are equal to one and the remaining $l>m$ satisfy $\abs{\lambda_l}< 1$, implying no oscillating fixed points. Indeed, employing that any eigenvalue with absolute value one is a root of unity, i.e. it has the form $\exp(i\frac{2\pi k}{p})$ with $k$ and $p$ integer~\cite{Perez-Garcia2007}, we could redefine the generation of the QGA as the $p$-composition $\mathcal{T}^p$ of the original channel when studying the asymptotic regime. In any case, this situation is not expected when employing the aforementioned subroutines, since the algorithm converges by construction.  

Given a decomposition of the initial state $\rho_{\text{in}} = \sum_l \omega_l W_l$ in terms of the eigenvectors from Eq.~\ref{eq:eigenvalueeq}, we define $\Lambda = \sum_{l=1}^m \omega_l W_l$ as the projection of $\rho_{\text{in}}$ into the convex set of fixed-points. As $\Lambda$ is a density matrix, the state for the $G$th generation can be straightforwardly obtained as
\begin{eqnarray}
\rho(G; \rho_{\text{in}}) &=& \mathcal{T}^G(\rho_{\text{in}})= \Lambda + \sum_{l\geq m+1} \omega_l \abs{\lambda_l}^G e^{i G \arg{\lambda_l}} W_l \nonumber\\
 &=& \Lambda + \mathcal{O}(\abs{\lambda_{m+1}}^G)\text{.}
\end{eqnarray}
This shows that the convergence of the QGA is exponential with a rate given by the magnitude of the second greatest eigenvalue of $\mathcal{T}$, $\abs{\lambda_{m+1}}$, which is called spectral subradius.

For the particular case $m=1$, the fixed point $\Lambda$ is unique and the spectral subradius is $\abs{\lambda_2}$. Note, that this unique fixed point depends on the problem Hamiltonian via the sorting subroutine. In this case, the expected value of an observable $\theta$ after $G$ generations can be estimated by
\begin{equation}\label{eq:asymptotic_obs}
\expval{\theta}(G; \rho_{\text{in}}) = \tr[\theta \rho (G; \rho_{\text{in}})] = \theta_{\infty} + \Delta \abs{\lambda_2}^G + \mathcal{O}(\abs{\lambda_3}^G)\text{,}
\end{equation}
where $\theta_{\infty}=\tr[\theta \Lambda]$ and $\abs{\lambda_2}$ only depend on the problem Hamiltonian, whereas $\Delta = 2\Re(\omega_2 e^{i G \arg{\lambda_2}} \tr[\theta W_2])$ depends on the initial state and the generation, but its absolute value is upper bounded by a constant.

% M. M. Wolf, Quantum Channels and OperationsÑGuided Tour (Lecture notes, 2012).
%M. M. Wolf, Quantum channels & operations guided tour, https://www-m5.ma.tum.de/foswiki/pub/M5/Allgemeines/MichaelWolf/QChannelLecture.pdf (2012).
\section{Results}
\label{sec:results}

\begin{figure*}[!htb]
\centering
\includegraphics[scale=0.9]{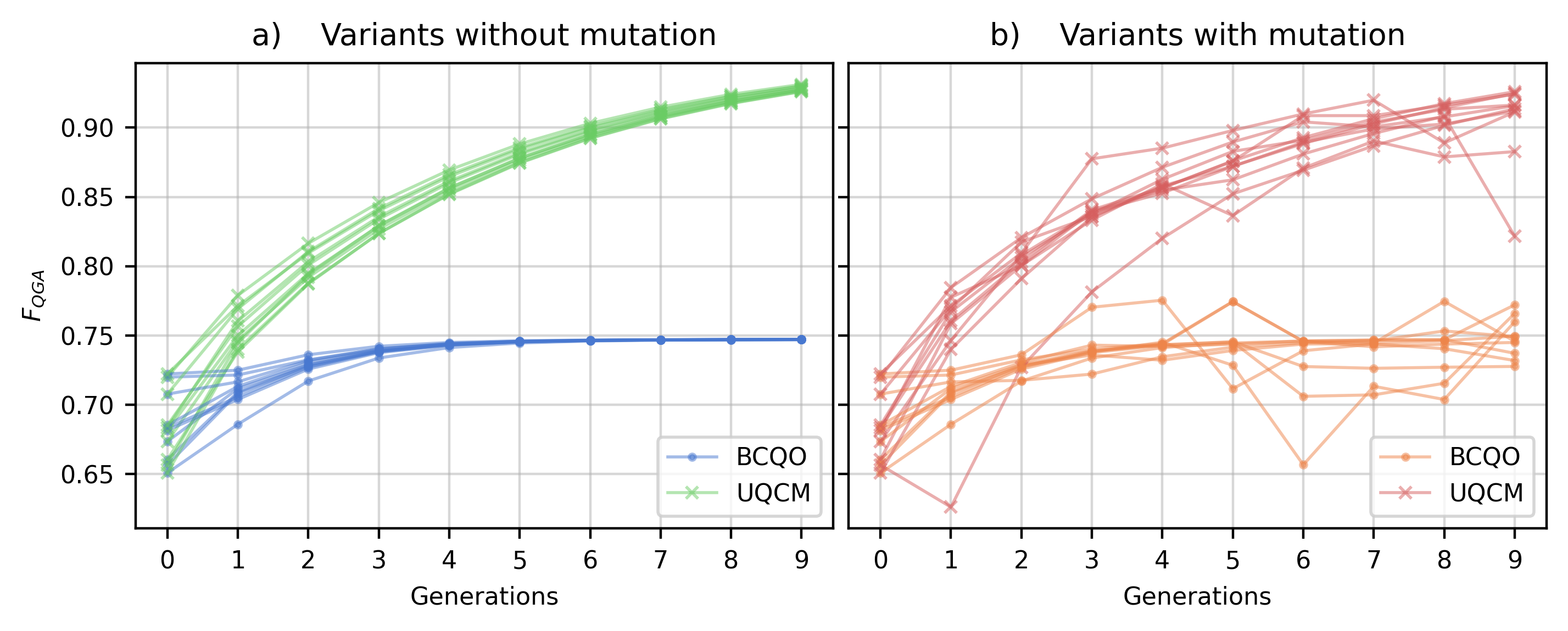}
\caption{Representative evolution of the fidelity between the best individual and the desired Hamiltonian ground state. We plot the evolution for the same $10$ initial populations considering four QGA variants applied on the same Hamiltonian. (a) The evolution for variants without mutation and (b) the evolution for variants with mutation, different markers represent variants with different cloning machines, namely, dots for BCQO variants and crosses for UQCM variants. Note that UQCM variants reach higher fidelity values, which is consistent with the statistical performance derived in Section~\ref{sec:convergence_and_performance}. The stochastic behavior of qubit tracing was considered accounting for the statistical information in the density matrix, whereas mutation was applied considering a different randomly chosen mutation pattern for each initial population and variant. This results in oscillations of the curves in (b), but we conclude an average convergence towards a stable fidelity value in Section~\ref{sec:convergence_and_performance}.}
\label{fig:convergence_curves}
\end{figure*}

In this section, we use two methods to benchmark four QGA variants with respect to a common set of problems. More specifically, for every problem, we simulate the evolution for each variant considering a random sample of initial states taken from a uniform Haar distribution. Additionally, we analyze the spectral properties of the quantum channel corresponding to the corresponding problem and variant on the grounds of Section~\ref{sec:fp_convergence}. The four variants consist of two different QCMs, namely BCQO and UQCM, together with the inclusion or not of the mutation subroutine. Due to computational constraints, the size of the simulated system is limited to $n=4$ individuals with chromosome length of $c=2$ qubits. Hence, the population is encoded in a total of $8$ qubits. We use a mutation probability of $p_m=\frac{1}{24}$, i.e. on average, one mutation in one of the $8$ qubits every $3$ generations.  Our benchmark consists of a sample of 200 problem Hamiltonians. Lastly, the figures of merit for this benchmark are the quantum fidelity between the best individual and the desired Hamiltonian ground state, and the corresponding convergence speed. 

Simulations were performed with matrix computations in Python-NumPy. We ran each simulation for $10$ generations, which were sufficient to estimate the figures of merit within the asymptotic regime due to the exponential convergence towards the convex set of fixed points, as explained in Section~\ref{sec:fp_convergence}. Additionally, we used for each problem Hamiltonian a random sample of $10$ initial quantum states taken from a uniform Haar distribution, which were the same for the four variants. This avoids the emergence of biases which can affect the performance of the cloning subroutine. We have numerically observed that the figures of merit merge with as few as $10$ sample states, thus this is statistically sufficient for our system size and leads to a good balance between precision and computational cost of the simulation. The mathematical reason behind this, as we will see, is that the algorithm has a unique fixed point. For each case, we recorded the problem Hamiltonian, the set of initial populations, the set of final populations, and the statistics of each individual in the Hamiltonian basis.

The quantum channel analysis was performed in Matlab. For this analysis, we have discarded the mutation subroutine, since we have observed in previous simulations a negligible effect in the performance while leading to a substantial increase in the computational cost. Firstly, we represent the quantum channel eigenvalue equation Eq.~\ref{eq:eigenvalueeq} in matrix form for both BCQO and UQCM variants applied to each problem Hamiltonian. To achieve it, we employ the mapping $\sum_k E_k \rho E_k^\dagger \rightarrow (\sum_k E_k \otimes E_k^*)|\rho\rangle$. Afterwards, we computed the six greatest eigenvalues and their respective eigenvectors, corresponding the fixed points to the ones with absolute value $1$. Remarkably, all cases and variants studied in this article show a unique fixed point. We expect that this situation is generic, but we leave a formal proof for further research. 

\subsection{Selection of the problem Hamiltonians}
\label{sec:selct_Hps}

%Any problem Hamiltonian can be described by its eigenstates and eigenvalues. However, consider two non-degenerate problem Hamiltonians, $H_P$ and $H_P'$, with different problem basis, $\ket{u_k}$ and $\ket{u_k'}$, related by the unitary transformation $U=\sum_{k=1}^{2^c} \ketbra{u_k'}{u_k}$, in Appendix~\ref{app:eq_nondegHp} we show that both selection subroutines are related by
%\begin{equation}
%\mathcal{T}_{S}' = U^{\otimes n} \mathcal{T}_S (U^{\dag})^{\otimes n}.
%\end{equation}
%This implies that the mathematical representation of the QGA for any non-degenerate Hamiltonian only depends on its problem basis and not the particular value of its energies.

Let us note that the problem Hamiltonian exclusively plays a role in the selection subroutine, and that this subroutine is only sensitive to ordinal position of the eigenvalues and not to their exact value. Since the states of problem basis defined in Section~\ref{sec:qselection} are sorted in increasing order according to their energy, the effect of the Hamiltonian can be fully described by its problem basis. In particular, it can be shown that given two problem Hamiltonians $H_P$ and $H_P'$ with bases $(\ket{u_1}, \dots, \ket{u_{2^c}})$ and $(\ket{u_1}', \dots, \ket{u_{2^c}'})$ related by $U=\sum_{k=1}^{2^c} \ketbra{u_k'}{u_k}$, their respective sorting subroutines are linked by a basis transformation on their Kraus operators
\begin{equation}
A_{\kappa}' = U^{\otimes n} A_{\kappa} (U^{\otimes n})^{\dag}\text{.}
\end{equation}

Therefore, we can map the task of generating random problem Hamiltonians to the task of generating random unitary transformations. We define the computational Hamiltonian $H_C$ as a diagonal Hamiltonian in the computational basis $\ket{u_k}=\ket{k}$ in canonical order, choosing without loss of generality eigenvalues proportional to their corresponding indexes, $\epsilon_k \propto k$. Afterwards, we sample a random $U$ from a uniform set of unitary operations, in order to generate the different problem Hamiltonians as $U H_C U^{\dag}$. 

\subsection{Figures of merit}
\label{sec:figures_of_meri}

We define the QGA fidelity after $G$ generations for the initial state of the population $\rho_\text{in}$ as the quantum fidelity between the state of the best individual found after $G$ iterations and the exact ground state of the Hamiltonian $\ket{u_1}$,
\begin{equation}
F_{\text{QGA}}(G;\rho_{\text{in}}) = \expval{\tr_{1\perp}[\rho(G; \rho_{\text{in}})]}{u_1}\text{,}
\end{equation}
where $\tr_{1\perp}$ is the partial trace operation over all the population but the best individual, and $\rho (G; \rho_{\text{in}})$ is the state of the population at the $G$th generation. Note that $F_{\text{QGA}}$ is computed after the sorting operation, since the best individual is located in the first register. Additionally, $F_{\text{QGA}}$ also corresponds to the probability of measuring the ground state energy.

The quantum fidelity is not a suitable metric for large systems, as it rapidly tends to zero for relatively small differences \cite{Girolami2021}. In fact, in the thermodynamic limit, two states are only either the same or orthogonal. Then, the expected energy of the best individual turns out to be a better figure of merit. This quantity additionally provides an adequate method for comparing this algorithm with other optimization approaches, and it is an adequate fitness function to evaluate the performance in a realistic scenario in which the exact solution of the problem is not available. Nevertheless, the expected energy of the best individual can be altered by the particular choice of the eigenvalues, thus the QGA fidelity is a better choice for the benchmarking analysis pursued in this article.

As $\ket{u_1}$ is a pure state, $F_{\text{QGA}}$ is the expected value of an observable. Additionally, we have empirically observed that the algorithm has generically a unique fixed point. This allows us to estimate the evolution of the QGA fidelity according to Eq.~\ref{eq:asymptotic_obs},
\begin{equation}
\label{eq:asymptotic_fidelity}
F_{\text{QGA}}(G;\rho_{\text{in}}) \approx F_{\infty} + \beta_{\text{in}} \gamma^G\text{,}
\end{equation}
where $F_{\infty}$ is the expected value of $F_{\text{QGA}}$ in the fixed point and $\gamma$ describes the convergence rate. When the fixed point is unique, both are independent from $\rho_{\text{in}}$, whereas $\beta_{\text{in}}$ is a bounded parameter depending on the initial state. In our analysis, we employ $F_{\infty}$ and $\gamma$ as figures of merit characterizing the quality of the final population and the convergence velocity, respectively.

\subsection{Quantum channel analysis and numerical simulations}
\label{sec:qch_vs_sim}

We have analyzed the accuracy and convergence speed of the algorithm in terms of the fidelity between the best individual and the desired ground state. These quantities are characterized by the asymptotic value $F_{\infty}$ and convergence-rate $\gamma$ introduced in Eq.~\ref{eq:asymptotic_fidelity}. We employed two methods to estimate their value: 1) fitting the parameters in Eq.~\ref{eq:asymptotic_fidelity} from the data obtained by the simulations described in the introduction of Section~\ref{sec:results}; and 2) computing them from the fixed points and spectral subradius corresponding to Eq.~\ref{eq:asymptotic_obs}. In this section, we show the agreement between both approaches.

In the first method, we fit the parameters in Eq.~\ref{eq:asymptotic_fidelity} from the data points $F_{\text{QGA}}(G;\rho_{\text{in}})$ by least-squares method. This is performed for every problem Hamiltonian and initial population. Then, $F_{\infty}$ and $\gamma$ values are averaged over different initial populations to estimate a single value for each problem Hamiltonian and avoid biases due to the initial state. We consider a burn-in period to ensure achieving the asymptotic regime, thus only values after four generations were considered in the fit.

In the second method, we compute both the fixed point and the spectral subradius of the channel by diagonalizing its matrix form. This allows us to predict $F_{\infty}=\expval{\tr_{1\perp}[\Lambda]}{u_1}$ and $\gamma = \abs{\lambda_2}$. As discussed in the introduction of Section~\ref{sec:results}, we have only studied the variants without the mutation subroutine with this method, since they have a negligible effect in the performance while leading to a substantial increase in the computational cost. It is important to highlight that all the channels analyzed in this article show a unique fixed point. We expect this property to be generic, i.e. for a randomly chosen problem Hamiltonian and any variant considered in this article this feature holds. However, there are corner cases in which there is a non-trivial convex set of fixed points. For instance, in Ref.~\cite{Ibarrondo2021} a fourth-fold degenerate example is constructed for a problem Hamiltonian diagonal in the computational basis employing the BCQO variant without mutation. Other situation in which this might happen is when the ground state of the problem Hamiltonian is degenerate. However, we leave a complete characterization of this scenario for further research.

The comparison between these two methods is shown in Figure~\ref{fig:qch_vs_sim}, where each point represents a different problem Hamiltonian. The vertical axis represents the $F_{\infty}$ and $\gamma$ obtained by fitting the simulation data, while the horizontal axis represents the same parameters obtained via quantum channel analysis. Fig.~\ref{fig:qch_vs_sim_a}(a) and Fig.~\ref{fig:qch_vs_sim_b}(b) depict the values of $F_{\infty}$ obtained respectively for BCQO and UQCM variants, both without mutation. Fig.~\ref{fig:qch_vs_sim_c}(c) and Fig.~\ref{fig:qch_vs_sim_d}(d) depict the values of $\gamma$ obtained respectively for BCQO and UQCM variants, again without mutation. The $R^2$ between simulation and quantum channel analysis for $F_{\infty}$ are $0.999$ for BCQO and $0.986$ UQCM.  Similarly, the $R^2$ between simulation and quantum channel analysis for $\gamma$ are $0.721$ BCQO and $0.999$ UQCM.  These correlation coefficients show a strong accordance between the results obtained with both methods, establishing the quantum channel analysis as a powerful tool to formally prove exponential convergence, as well as providing mathematical techniques to show bounds for the accuracy. 

\subsection{Convergence and performance}
\label{sec:convergence_and_performance}

\begin{figure*}[!htb]
\centering
\includegraphics[scale=0.8]{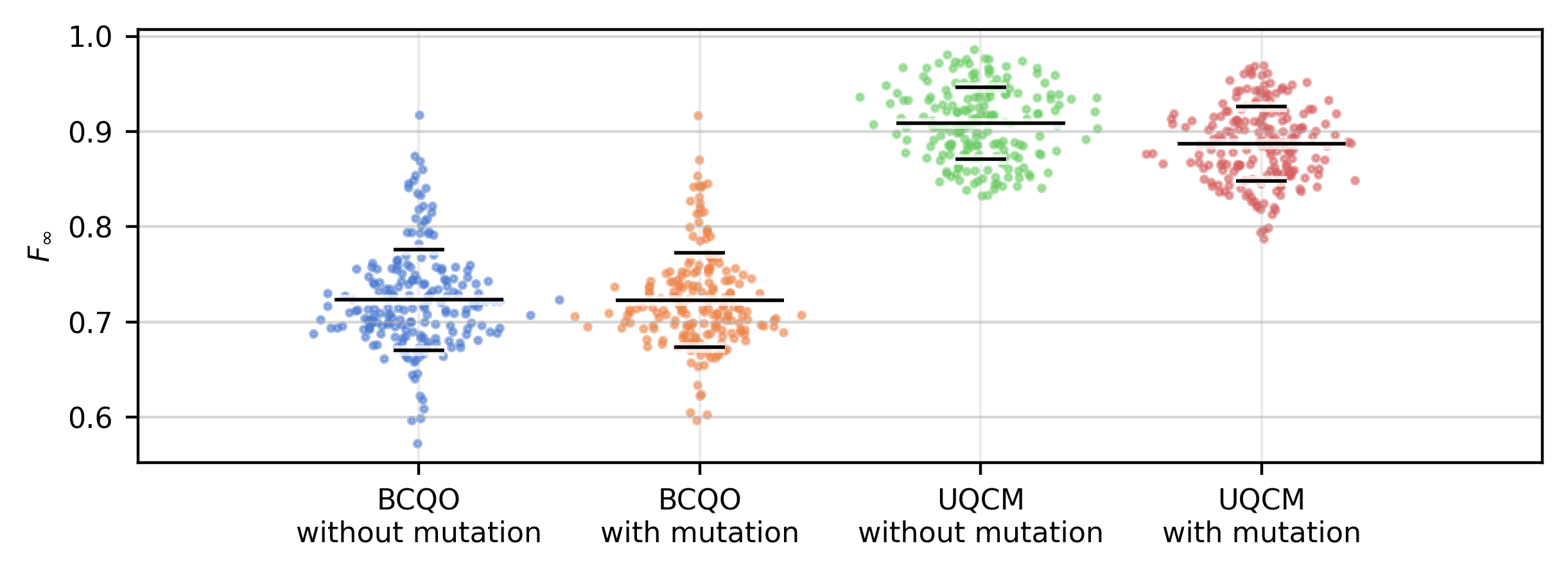}
\caption{Final fidelity between the best individual and the desired Hamiltonian ground state, $F_{\infty}$. We applied four variants -- employing two different cloning machines, BCQO or UQCM, and excluding or including mutation -- to a sample of randomly generated problem Hamiltonians. The vertical position of each point is the average value from 10 initial populations for each problem Hamiltonian, and the horizontal position is jittered to reduce the visual overlap between points. BCQO variants hold the same mean and standard deviation -- $0.72$ and $0.05$ -- independently on the presence or absence of mutation. The UQCM variant without mutation has mean and standard deviation equal to $0.91$ and $0.04$, whereas, for the UQCM variant with mutation are $0.89$ and $0.04$.}
\label{fig:variant_cmp}
\end{figure*}

In Figure~\ref{fig:convergence_curves}, we show a representative evolution of the $F_{\text{QGA}}$ for each variant, illustrating the exponential convergence modeled in Eq.~\ref{eq:asymptotic_fidelity}. Recall that we used identical initial populations for the four variants. Each initial population and variant is represented by a line, and all lines correspond to the same problem Hamiltonian. Cases without mutation in Fig~\ref{fig:convergence_curves}(a) showcase a better fit than Fig~\ref{fig:convergence_curves}(b) because mutation was applied as a unitary gate despite being stochastic, producing random patterns in the fidelity. Note that in the rest of the subroutines stochastic elements are integrated in the density matrix description.

According to the numerical fit, the BCQO variant without mutation shows $\gamma$ values between $0.18$ and $0.88$, with an average of $0.50$ and a standard deviation of $0.13$. Meanwhile, the UQCM variant without mutation shows $\gamma$ values between $0.47$ and $0.79$, with an average of $0.60$ and a standard deviation of $0.07$. Additionally, we compared the variants for each problem Hamiltonian and obtained that the BCQO variant tends to lower $\gamma$ values in $77\%$ of cases, with a $95\%$ confidence interval of $\pm 6$. Therefore, the BCQO variant presents a faster convergence on average. 

Figure~\ref{fig:variant_cmp} summarizes the performance of the variants in terms of the fidelity $F_{\infty}$ between their best individual and the ground state. We can note that the success probability is above 0.6 for virtually every problem Hamiltonian. Remarkably, BCQO variants hold the same mean and standard deviation -- $0.72$ and $0.05$ -- regardless of the presence or absence of mutation. The mean and standard deviation of the UQCM variant without mutation are $0.91$ and $0.04$ respectively, whereas those of the UQCM variant with mutation are $0.89$ and $0.04$ respectively. As we can see, the UQCM variant without mutation outperforms the others in the studied cases, yielding $25\%$ higher fidelity than BCQO ones and $2\%$ higher fidelity than the UQCM variant with mutation. Hence, replicating the individuals with the UQCM produces generally better results in terms of the fidelity with the desired Hamiltonian ground state. Note, however, that this result is obtained on a set of randomly generated Hamiltonians and that BCQO variants may improve their results for problem Hamiltonians that nearly commute with the quantum observable chosen to define the BCQO.

Overall, we observe that both $\gamma$ and $F_{\infty}$ vary more with respect to the problem Hamiltonian for the BCQO variants. This is because the cloning fidelity of BCQO strongly depends on the input state, and whether it is diagonal to the basis of the cloning observable, unlike UQCM. Regarding convergence rate, entanglement could play a role, as stronger entanglement in the cloning process leads to a larger loss of information in the selection process. We have experimentally observed that BCQO cloning produces stronger entanglement than UQCM, which could intuitively explain its faster convergence. This can also be observed on a quantum-channel level, where the spectral decomposition of the reset and cloning subroutines are more extreme for BCQO than for UQCM. Indeed, the eigenvalues of the former are always either $1$ or $0$, originating a more abrupt collapse. Note that this does not imply a better minima, just a faster convergence.

The employed method for the mutation subroutine does not introduce any meaningful performance improvement. It produced a marginal effect in the BCQO variant, but also a slight performance decrease in the UQCM variant. Further research is required in order to understand the role of mutation and the use of other approaches. However, for the relatively small search space that we have explored in our analysis, exploitation takes precedence over exploration, which reduces the importance of this subroutine.

\section{Conclusions} 
\label{sec:conclusions}

Here, we have introduced a QGA comprising the fundamental elements which characterize GAs. This was attained by codifying the individuals in non-orthogonal quantum states supported in independent registers, a distinctive feature with respect to previous approaches. Moreover, we have codified the optimization problem in a non-diagonal Hamiltonian and the cost function in the energy of the individual with respect to this Hamiltonian. The algorithm has a modular structure composed of quantum selection, crossover, and mutation subroutines. At the cost of introducing some ancillary qubits, the selection was performed as a reversible sorting algorithm with respect to the problem Hamiltonian, tracing out the lowest ranked individuals. We carried out the replication step in the crossover via a partial quantum cloning machine and the combination of half of the genome by swapping the corresponding qubits. We have benchmarked two paradigmatic quantum approximated cloning machines: biomimetic cloning of quantum observables and Bu\v zek-Hillery universal quantum cloning machine. We have generated a sample of $200$ random problem Hamiltonians, ran the quantum algorithm for $10$ generations, and compared for both cloning machines the convergence ratio and fidelity of their corresponding solutions with respect to the real ground state. Then, this numerical analysis showed that the convergence speed employing the BCQO is larger than with UQCM in $77$\% of the cases. However, when we focus on the fidelity of their best individual with respect to the ground state, we observed that UQCM always outperforms BCQO in the studied cases with an average improvement in the fidelity of $25$\%. Lastly, we concluded that introducing mutations, implemented by means of randomly allocated Pauli gates, had a negligible effect on the fidelity of the best individual. In fact, even though there were small changes case by case, both the mean and the standard deviation were identical with and without mutation for the BCQO and a slight advantage of $2$\% in the absence of mutations for the UQCM. Finally, we have expressed our subroutines as quantum channels, such that each generation of the algorithm, which is also a quantum channel, is a composition of them. It follows that the iteration of the algorithm corresponds to the composition of this channel with itself. Therefore, we can employ the spectral theory of quantum channels to formally prove an exponential convergence of the algorithm towards the fixed point of the channel, as well as to bound its convergence rate by its spectral subradius. Remarkably, both this prediction and the final quantum state accurately match our numerical simulations. Indeed, the correlation coefficients between the fidelities obtained by means of numerical simulations and quantum channel techniques were $R^2=0.999$ for BCQO and $R^2=0.986$ for UQCM. Similarly, the correlation coefficients between predicted and numerically obtained convergence rates were $R^2=0.721$ for BCQO and $R^2=0.999$ for UQCM. This approach can be extended to other non-unitary iteration-based quantum algorithms.

\section*{Acknowledgements}

The authors acknowledge financial support from Spanish Government PGC2018-095113-B-I00 (MCIU/AEI/FEDER, UE), Basque Government IT986-16, Spanish Ram\'on y Cajal Grant RYC-2020-030503-I and the QUANTEK project from ELKARTEK program (KK-2021/00070), as well as from QMiCS (820505) and OpenSuperQ (820363) of the EU Flagship on Quantum Technologies, and the EU FET-Open projects Quromorphic (828826) and EPIQUS (899368).

\bibliographystyle{IEEEtran}
%\bibliography{literatureQGA}

\end{document}